\documentclass[prl,floatfix,twocolumn,amsmath]{revtex4}
\usepackage{graphicx}
\usepackage{subfigure}
\usepackage{color}
\usepackage{multirow}

\begin{document}

\def\ket#1{|#1\rangle}
\def\bra#1{\langle#1|}
\def\av#1{\langle#1\rangle}
\def\myarrow{\mathop{\longrightarrow}}
\def\ua{\uparrow}
\def\da{\downarrow}
\setlength\abovedisplayskip{9pt}
\setlength\belowdisplayskip{9pt}
\setlength\belowcaptionskip{-8pt}

\title{Precision measurement of transverse velocity distribution of a strontium atomic beam}

\author{F. Gao$^{1,2}$}
\author{H. Liu$^{1,2}$}
\author{P. Xu$^1$}
\author{X. Tian$^{1,2}$ }
\author{Y. Wang$^1$}
\author{J. Ren$^1$}
\author{Haibin Wu$^{3}$}
\email{hbwu@phy.ecnu.edu.cn}
\author{Hong Chang$^{1,3}$}
\email{changhong@ntsc.ac.cn}

\affiliation{$^{1}$ CAS Key Laboratory of Time and Frequency Primary Standards, National Time Service Center, Xi'an 710600, China}
\affiliation{$^{2}$ University of Chinese Academy of Sciences, Beijing 100049, China}
\affiliation{$^{3}$ State Key Laboratory of Precision Spectroscopy, Department of Physics, East China Normal University, Shanghai 200062, China}

\date{\today}

\begin{abstract}
 We measure the transverse velocity distribution in a thermal Sr atomic beam precisely by velocity-selective saturated fluorescence spectroscopy. The use of an ultrastable laser system and the narrow intercombination transition line of Sr atoms mean that the resolution of the measured velocity can reach 0.13 m/s, corresponding to 90$\mu K$ in energy units. The experimental results are in very good agreement with the results of theoretical calculations. Based on the spectroscopic techniques used here, the absolute frequency of the intercombination transition of $^{88}$Sr is measured using an optical-frequency comb generator referenced to the SI second through an H maser, and is given as 434 829 121 318(10) kHz.
\end{abstract}
\maketitle

Atomic (or molecular) beams, which have now become standard laboratory tools, play very important roles in the field of atomic, molecular and optical physics. These beams have been widely used in the determination of atomic structures~\cite{atomstructure}, measurement of physical constants~\cite{physicsconstant}, studies of chemical reactions~\cite{reaction} and atomic frequency standards~\cite{frequencystandard}. In all these applications, measurement of the velocity distribution of the atomic beams is both necessary and highly important. Many spectroscopic techniques have been developed to perform these measurements~\cite{beam1,beam2,beam3,beam4,beam5,beam6,beam7,beam8}. However, the precise determination of the velocity distribution of an atomic beam, and especially that of a well collimated atomic beam, remains challenging because of the low signal-to-noise ratio and relatively low spectroscopy resolution of these beams.

In this paper, we report the precise measurement of the transverse velocity distribution of a well-collimated thermal strontium beam by velocity-selective saturated fluorescence spectroscopy. The measurement accuracy of the transverse velocity of the atomic beam is greatly increased by the combination of an ultrastable laser system and high-resolution spectroscopy of the intercombination transition of $^{88}$Sr. The measurement resolution for the velocity can reach 0.13 m/s, corresponding to 90 $\mu K$ in energy units. Because of the confinement of the cylinder nozzle of the atomic beam, the transverse velocity distribution is no longer the well-known Maxwell-Boltzmann distribution. The measured results show good agreement with the theoretically predicted results~\cite{theory}. Using the spectroscopic techniques developed here, the absolute frequencies of the isotopes of strontium atoms are measured using an optical frequency comb generator referenced to the SI second through an H maser. Particular attention is paid to the intercombination transition of $^{88}$Sr. The frequency is measured as 434 829 121 318(10) kHz.

\begin{figure}[htbp]
\centerline{\includegraphics[width=.85\columnwidth]{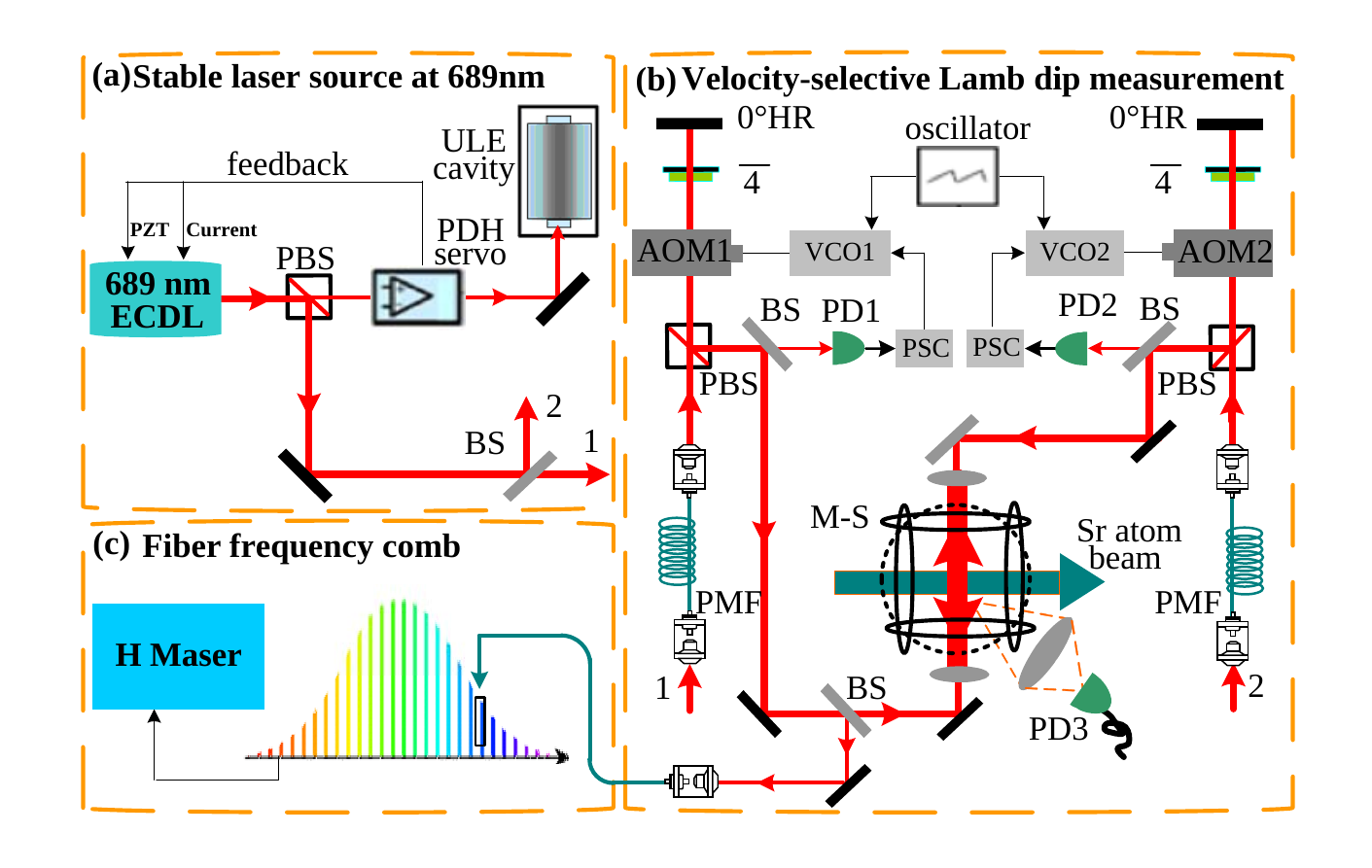}}
\caption{Experimental setup. (a) The narrowed linewidth of the laser system (689 nm). PBS: polarized beam splitter, ULE: ultra-low expansion optical cavity, PDH: Pound-Drever-Hall frequency locking system. (b) Experimental setup used for velocity-selective saturated fluorescence spectroscopy. The frequencies of the two counterpropagating laser beams are scanned using a saw-tooth waveform at 0.1 Hz. PMF: single-mode polarization-maintaining fiber, AOM: acousto-optic modulator, VCO: voltage-controlled oscillator, PSC: power stabilization servo controller, M-S: Earth magnetic shield. (c) Optical frequency measurement using a fiber frequency comb (Menlo FC1500), with respect to an H maser at a radio frequency signal of 10 MHz.  }
\end{figure}

Velocity-selective saturation (fluorescence) spectroscopy is a powerful technique that is widely used for high resolution measurements. We use it here to measure the velocity distribution of the atomic beam precisely. The measurement principle is as follows. Consider two-level atoms interacting with the counterpropagating probe beam and pump beam; the probability of the atoms being in excited states is proportional to the velocity of the atoms, the light intensities and the scattering rate of the excited state $\gamma$. When the probe and pump beams have the same frequency, the group of atoms with zero velocity see the light beams at resonance and show a Lamb dip at the atomic resonant frequency with a large Doppler background. The peak amplitude can be used to determine the number of atoms. When the frequencies of the pump and probe lasers are different, only the atoms with velocity $v=\Delta\omega/(2k)$ see the light beams at resonance, where $\Delta\omega$ is the frequency difference between the beams and $k$ is the wave vector of the laser light. The spectroscopic resolution of this method is limited by the natural linewidth of the excited state, and therefore the velocity resolution is $v=(\gamma/f) c$, where $f$ and $c$ are the resonant frequency and the light speed in a vacuum, respectively. The atom used here is strontium, which has four stable natural isotopes, comprising bosonic $^{88}$Sr (82.58\%), $^{86}$Sr (9.86\%), $^{84}$Sr (0.56\%), with nuclear spin I=0, and fermionic $^{87}$Sr (7.0\%), with I=9/2. The intercombination line $5^1S_0-5^3P_1$, because of its high frequency fraction ($f/\gamma$), has been widely studied in optical frequency standards~\cite{frequency1, frequency3, frequency4}. In this paper, we use this transition to precisely measure the velocity distribution of the atomic beam.


The experimental setup is shown in Fig. 1. A commercial extended cavity diode laser (ECDL) (Topical-110) is used as the spectroscopy laser, and can typically deliver 16 mW at 689 nm. The laser linewidth is reduced by locking the laser to an ultrastable optical cavity via the Pound-Drever-Hall scheme; the phase modulation is produced by a resonant electro-optic modulator (EOM), which is driven at 5 MHz. The cavity is made of an ultralow expansion glass with a finesse of 12000. To avoid a residual standing wave in the EOM, which induces spurious amplitude modulation (AM) on the locking signal, a 60 dB optical isolator is placed between the EOM and the cavity. The power of the locking beam is maintained at as low a level as possible to minimize the frequency shift caused by light heating and optical feedback. The linewidth and the fractional frequency drift are about 200 Hz and $2.8\times 10^{-13}$ at 1 s, respectively.

The strontium atomic beam is obtained from strontium metal heated to 823 K (550 $^o$C) in an oven. A bundle of stainless steel capillaries (Monel-400) are used to collimate the beam. The length and the internal diameter of the capillary are 8 mm and 200 $\mu m$, respectively. The residual atomic beam divergence is 25 mrad, and the typical atomic flux is estimated to be $10^{12}/s$. The vacuum for the atomic beam is maintained at $1\times 10^{-8}$Torr with a 40 l/s ion pump.

Velocity-selective saturation (fluorescence) spectroscopy is obtained by the use of two counterpropagating laser beams (i.e. the probe and pump beams) perpendicular to the atomic beam. These beams are generated by two acousto-optical modulators (AOMs) with the same double-pass configurations. The AOMs are controlled by the same oscillator but with different voltage-controlled oscillators (VCOs). A homemade circuit is used to prevent the oscillator frequencies from disturbing each other. The frequency resolution is approximately 0.1 Hz. The power is stabilized by a servo, which is better than $10^{-4}$. Both the pump and probe beams have been expanded to have 1.2 cm waists (1/$e^2$ diameter). The fluorescence signal is collected by a large diameter lens in a direction that is orthogonal to the atomic beam and the laser light beams on a high-sensitivity detector. A magnetic shield was placed in the interaction region to prevent the Zeeman effect being caused by the Earth’s magnetic field.

\begin{figure}[htbp]
\centerline{\includegraphics[width=.85\columnwidth]{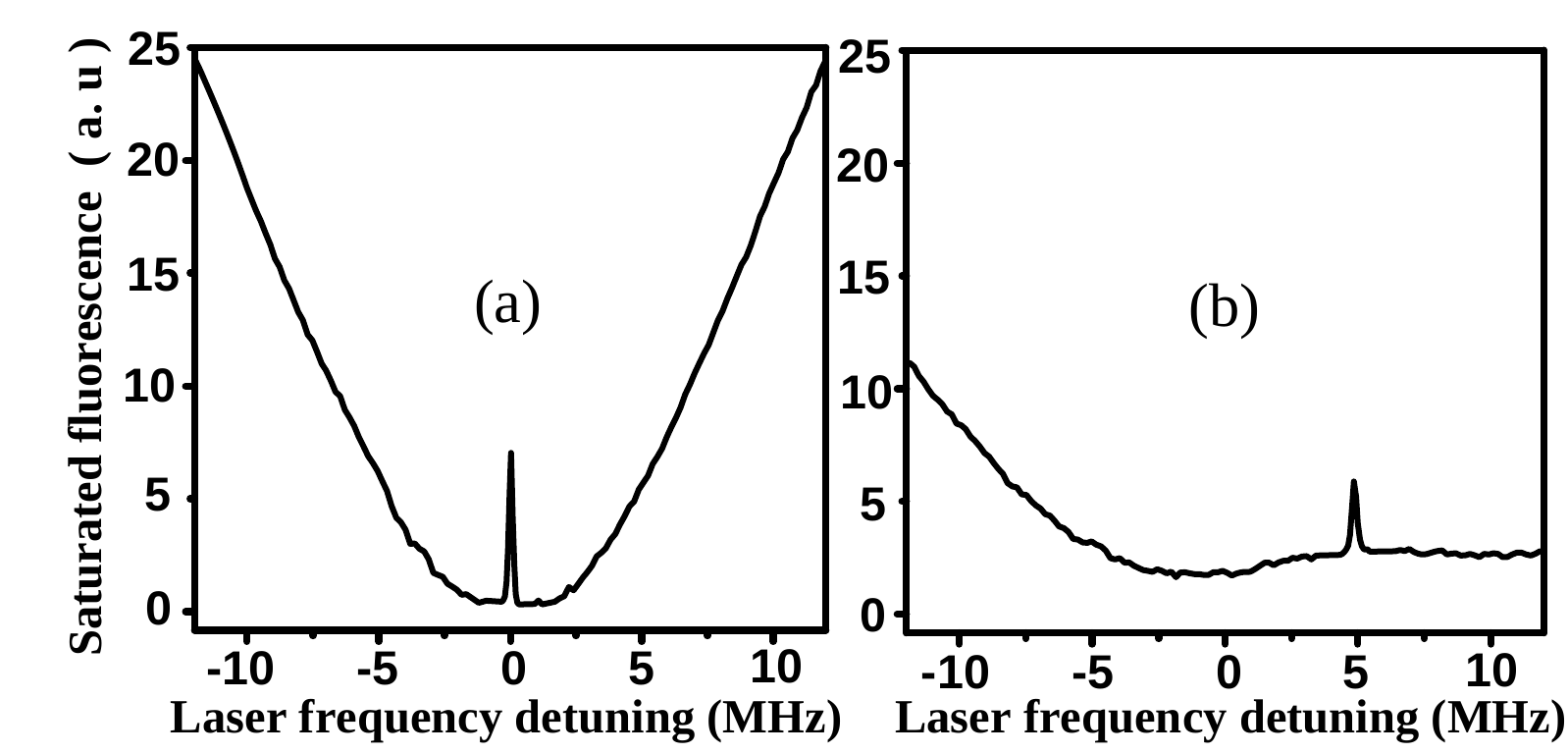}}
\caption{Velocity-selective saturated fluorescence spectra: (a) $\Delta=0 MHz$, corresponding to the condition where a group of atoms with velocity $v=0$ is measured; (b) $\Delta=5 MHz$, corresponding to the condition where a group of atoms with velocity $v=3.45 m/s$ is measured. }
\end{figure}

 By scanning the pump and probe light beams at different frequencies, we observe the velocity-selective saturated fluorescence spectra. Fig. 2 shows two typical spectra for $\Delta=0$ and $\Delta=5\,MHz$, where $\Delta$ is the frequency difference between the probe and pump light beams. The linewidth of the central peak in Fig. 2(a) is approximately 180 kHz, which is larger than its natural linewidth of 7.5 kHz. This effect mainly stems from power broadening. The beam intensity of 800 $\mu W/cm^2$ (the saturation intensity is 3 $\mu W/cm^2$) is chosen to produce a large signal-to-noise ratio for the spectroscopy signal. The contributions of second-order Doppler broadening (0.55 kHz) and transit time broadening (1.06 kHz) are small and can be neglected. With this linewidth for the spectroscopy, the transverse velocity of the magnitude at 0.13 $m/s$ (corresponding to 90 $\mu K$ in energy units) can be measured.

\begin{figure}[htbp]
\centerline{\includegraphics[width=.75\columnwidth]{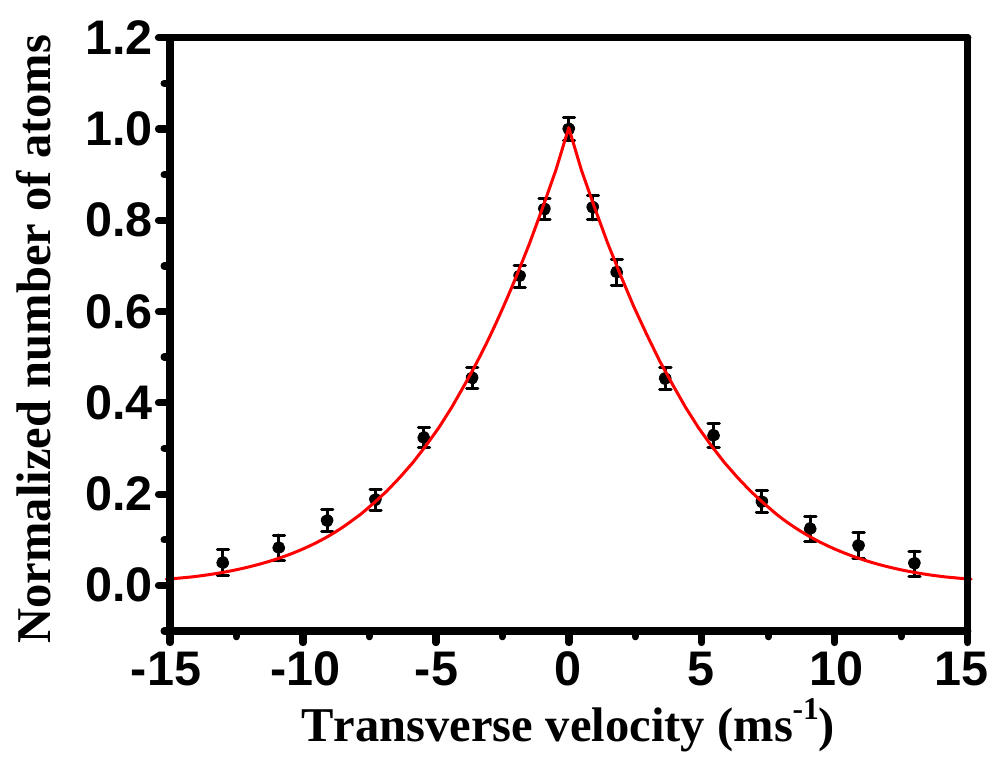}}
\caption{ The transverse velocity distribution of the atomic beam. The black dots represent the data for the experimental results. The error bars denote statistical fluctuations that arise from measurement uncertainty for the amplitudes of the signals. The red solid curve shows the results of the theoretical calculations. All parameters are taken from the experimental measurements. There are no free parameters. }
\end{figure}

For different frequency detunings of the pump and probe light beams, groups of atoms with different velocities see the light beams at resonance. The number of atoms can be measured from the amplitude of the saturated fluorescence spectroscopy signal, and therefore the velocity distribution can be obtained, and is shown in Fig. 3. The number of atoms has been normalized with respect to the number of atoms when $\Delta=0$. The maximum frequency scanning range for our AOMs is approximately $20\,MHz$, corresponding to $13. 78\, m/s$ for the maximum measurable transverse velocity of the atoms. Fig. 3 clearly shows that the transverse velocity distribution of the thermal atomic beam is no longer a Maxwell-Boltzmann-like distribution. The confinement of the capillaries leads to this new distribution, which was predicted in Ref.~\cite{theory} as
\begin{equation}
P(v,a)=\frac{|v|\exp(-v^2/v_0^2)\Gamma(-1/2,v^2L^2/v_0^2 a^2)}{\sqrt{8\pi v_0^2(1+a^2/L^2)^{1/2}}},
\end{equation}
where $v_0\equiv\sqrt{2k_BT/m}$ is the most probable velocity for the atoms, $k_B$ is the Boltzmann constant, T is the oven temperature, $m$ is the mass of the atoms, $a$ is the internal diameter of the collimator and $L$ is the length of the collimator. $\Gamma(-1/2,v^2L^2/v_0^2 a^2)$ is the incomplete Gamma function. Based on the experimental parameters, the results of the theoretical calculation of the velocity distribution are plotted as the solid curve (red curve) in Fig. 3. There are no free parameters. The theoretical curve clearly shows very good agreement with our measurement results. The influence of the three other isotopes, $^{86}$Sr, $^{87}$Sr and $^{84}$Sr, on the measurement results is small because of large frequency differences and the low natural abundance of these isotopes.

\begin{figure}[htbp]
\centerline{\includegraphics[width=.8\columnwidth]{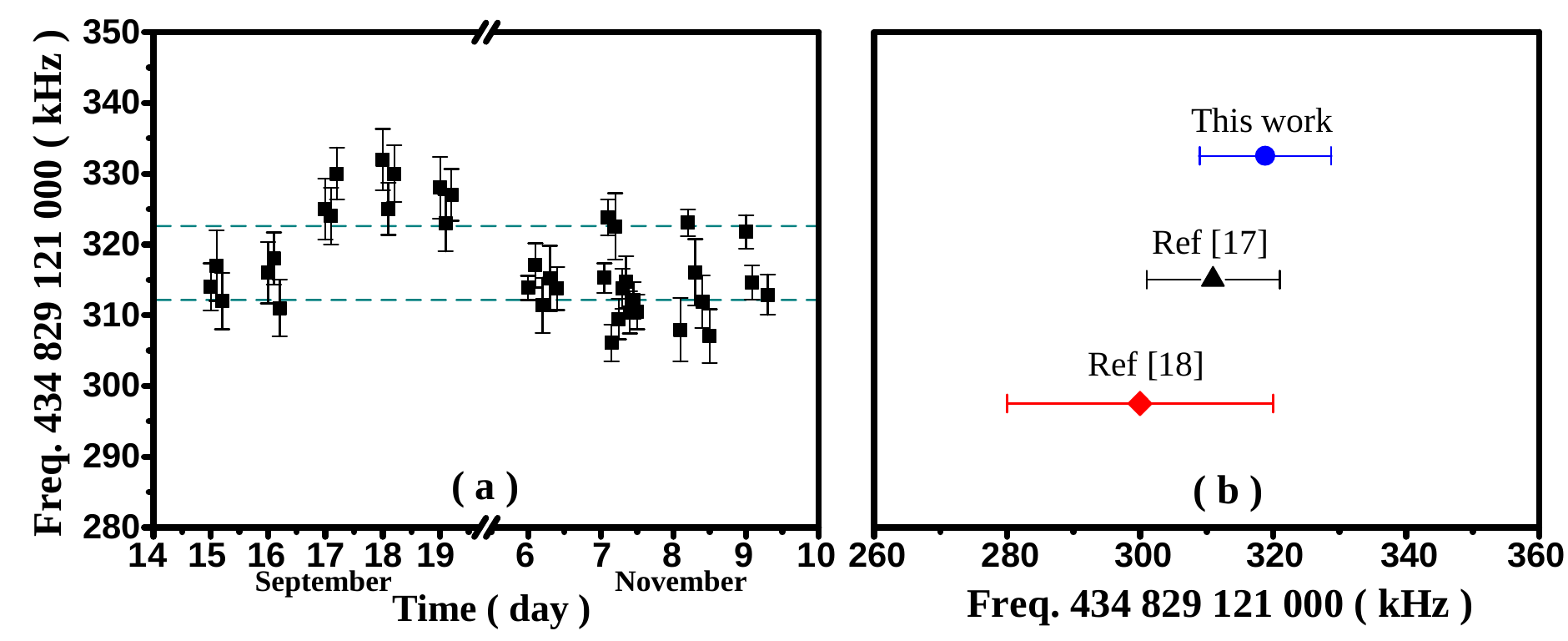}}
\caption{(a) Frequency measurement of the $^{88}Sr$ $5^1S_0\to 5^3P_1$ transition. The error bars correspond to the standard deviation for each data set. (b) Comparison of the optical frequency measurement of the intercombination line of $^{88}Sr$ with the results of previous measurements~Refs~\cite{mea1, mea2}. The black triangle represents the data from Ref~\cite{mea1}, the red square represents the data from Ref~\cite{mea2}, and the blue dot represents the data from our measurement. }
\end{figure}

\begin{table}[htbp]
\caption{
 \label{TabBiasUncertainty}
 Optical frequency measurement results and uncertainties for all four natural isotopes of strontium atoms.\\\\}
    \begin{tabular}{| l | l | l | }
     \hline
     \hline
     Isotopes & $5^1S_0 \to 5^3P_1$ & Frequency (kHz) \\ \hline
     $^{88}Sr$ & $ J=0$$\to J'=1$ & 434 829 121 318 (10) \\ \hline
      \multirow{4}{*}{$^{87}Sr$} & $F=9/2$$\to F'=7/2$ & 434 830 473 227 (45) \\
                                  & $F=9/2\to F'=9/2$ & 434 829 342 995 (55)\\
                                  & $F=9/2\to F'=11/2$& 434 827 879 835 (50) \\ \hline
     $^{86}Sr$ & $J=0$$\to J'=1$ & 434 828 957 500 (15) \\ \hline
     $^{84}Sr$ & $J=0$$\to J'=1$ & 434 828 769 730 (100) \\ \hline
     \hline
 \end{tabular}
\end{table}

 We reduce the power of the pump and probe light beams and use our narrowed linewidth laser (200 Hz), and the saturation fluorescence spectroscopy linewidth is approximately 55 kHz. The optical frequency of the Sr atoms is measured using a commercial optical-frequency comb (Menlo FC1500). The repetition rate and the carrier offset envelope frequency are locked to an H maser. Fig. 4(a) shows the results of measurements of the $^{88}$Sr transition frequency taken over a period of several days. The error bars correspond to the
standard deviation. The absolute frequency of the intercombination transition of $^{88}$Sr is 434 829 121 318(10) kHz. A comparison with the results of previous measurements in Ref~\cite{mea1, mea2} is presented in Fig. 4(b), and shows reasonable agreement. The optical frequencies of $^{86}$Sr, $^{87}$Sr and $^{84}$Sr are measured using high laser intensities (2.1 $mW/cm^2$, 4.3 $mW/cm^2$, and 12 $mW/cm^2$, respectively); because of the low natural abundance of these isotopes, the high power is required to obtain a good signal-to-noise ratio. The measurement results for the optical frequencies of all four natural isotopes of atomic strontium are summarized in Table I.

In conclusion, we report on the precise measurement of the transverse velocity distribution of a well-collimated thermal Sr atomic beam using the velocity-selective saturation fluorescence spectroscopy technique. The combination of the ultrastable laser system and the narrow linewidth of the intercombination transition means that the velocity measurement resolution is greatly improved. The detectable minimum of the velocity is approximately $0.13\, m/s$ (corresponding to $90\,\mu K$ in energy units). The velocity distribution is no longer likely to be a normal Maxwell-Boltzmann distribution. Instead, it shows a counter-intuitive umbrella shape. The measured data are in very good agreement with the results of the theoretical calculations. We also measured the optical frequency of the $5^1S_0\to 5^3P_1$ transition through an optical-frequency comb generator referenced to the SI second via an H maser. The measured absolute frequency of the intercombination transition of $^{88}$Sr is comparable to the results of previous measurements. Accurate optical frequency values for all other isotopes are also presented, and these results may provide a benchmark for subsequent measurements.

This research is supported by the National Natural Science Foundation of China (Grant Nos. 11074252 and 61127901), and the Key Projects of the Chinese Academy of Sciences (Grant No. KJZD-EW-W02).

\end{document}